\documentclass[aps,prl,reprint,groupedaddress]{revtex4-1}
\usepackage{dcolumn}
\usepackage{graphicx,amsmath,amssymb,braket,siunitx}
\DeclareSIUnit\gauss{G}
\DeclareSIUnit\dbm{dBm}
\DeclareSIUnit\sccm{sccm}
\DeclareSIUnit\gamma{\mbox{$\Gamma$}}
\DeclareSIUnit\sig{\mbox{$\sigma$}}

\begin{document}

\title{Magnetic trapping and coherent control of laser-cooled molecules}
\author{H. J. Williams, L. Caldwell, N. J. Fitch, S. Truppe, J. Rodewald, E. A. Hinds, B. E. Sauer, M. R. Tarbutt}
\email[]{m.tarbutt@imperial.ac.uk}
\affiliation{Centre for Cold Matter, Blackett Laboratory, Imperial College London, Prince Consort Road, London SW7 2AZ UK
}

\begin{abstract}
We demonstrate coherent microwave control of the rotational, hyperfine and Zeeman states of ultracold CaF molecules, and the magnetic trapping of these molecules in a single, selectable quantum state. We trap about \SI{5e3} molecules for \SI{2}{\second} at a temperature of \SI{65+-11}{\micro\kelvin} and a density of \SI{1.2e5}{\per\cubic\centi\metre}. We measure the state-specific loss rate due to collisions with background helium.    
\end{abstract}

\pacs{}
\maketitle

Techniques for producing and controlling ultracold molecules are advancing rapidly, motivated by a wide range of applications. These include precise measurements that test the foundations of theoretical physics~\cite{Safronova2017,DeMille2017}, quantum state resolved collisions and chemistry~\cite{Krems2008}, quantum computation~\cite{DeMille2002,Yelin2006,Andre2006} and simulation~\cite{Micheli2006,Barredo2016} and the study of dipolar quantum gases~\cite{Lahaye2009}. These applications call for molecules in a single selectable quantum state, trapped for long periods and exhibiting long coherence times. Some demand high phase-space density, and most require coherent control over the rotational and hyperfine states. Such control has recently been achieved~\cite{Yan2013, Will2016,Gregory2016,Park2017} for heteronuclear bialkali molecules produced by the association of ultracold atoms~\cite{Ni2008,Takekoshi2014,Molony2014,Park2015,Guo2016}. Great efforts are also being made to cool molecules directly, for example by optoelectrical Sisyphus cooling~\cite{Prehn2016}, or by direct laser cooling and magneto-optical trapping~\cite{Barry2014,McCarron2015,Steinecker2016,Truppe2017b,Anderegg2017, Williams2017}. These methods can produce ultracold molecules with greater chemical diversity and with large electric and magnetic dipoles, as is often desired. The magneto-optical trap (MOT) is an excellent tool for collecting and cooling molecules, and promises to be the starting point for many applications of ultracold molecules, just as it has been for ultracold atoms. However, it does not allow quantum state control, provides limited phase-space density, and has a limited lifetime due to optical pumping into states not addressed by the lasers. Thus, molecules must be transferred out of the MOT and into a conservative trap where the lifetime can be long, the quantum state can be selected and preserved, and the phase-space density can be increased, for example by sympathetic, evaporative or Raman sideband cooling. Magnetic trapping has been crucial for exploiting ultracold atoms, and magnetic traps have previously been used to confine molecules produced at $\sim\SI{100}{\milli\kelvin}$ by buffer-gas cooling, Stark deceleration and Zeeman deceleration~\cite{Weinstein1998,Sawyer2007, Hogan2008, Tsikata2010,Riedel2011,Lu2014,Akerman2017}. Here, we demonstrate coherent control and magnetic trapping of laser-cooled molecules, which are key steps towards the applications discussed above. Starting from a MOT of CaF~\cite{Williams2017}, we compress the cloud to increase its density, cool the molecules to sub-Doppler temperatures in an optical molasses~\cite{Truppe2017b}, optically pump them into a single internal state, transfer them coherently to a selectable rotational, hyperfine and Zeeman level, and confine them for seconds in a magnetic trap.

\begin{figure*}[!tb]
\includegraphics[width=\textwidth]{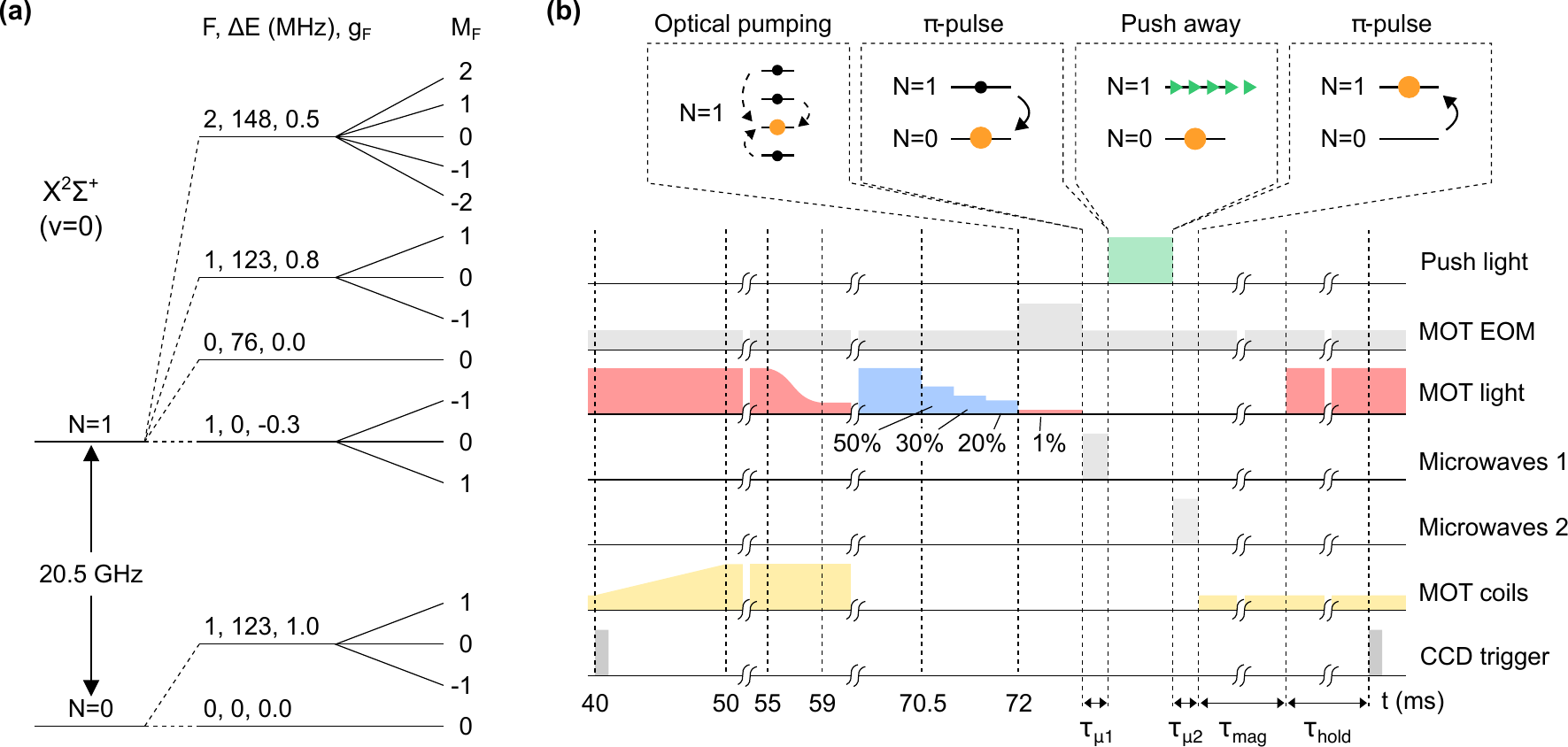}
\caption{(a) Energy levels (not to scale) of the two lowest-lying rotational states of the $X^{2}\Sigma^{+}(v=0)$ state of CaF. Hyperfine splittings and magnetic g-factors are given. (b) Timing sequence of the experiment (not to scale), starting from $t=\SI{40}{\milli\second}$ when the first fluorescence image is taken. The MOT light sequence is color coded according to the detuning of the light. The MOT is first compressed by ramping up $B'$. Then, the molecules are cooled in a blue-detuned molasses, optically pumped into $\ket{1,0,0}$, and transferred to $N=0$ with a microwave pulse. A push pulse removes those remaining in $N=1$, then (optionally) another microwave pulse transfers from $N=0$ back to $N=1$. Finally, the magnetic trap is turned on.}
\label{fig:structure_timing}
\end{figure*}

Our setup is the same as used previously~\cite{Truppe2017b,Williams2017}, with the addition of microwave components to drive the rotational transition. A pulse of CaF emitted at time $t=0$ from a cryogenic buffer gas source~\cite{Truppe2017c} is slowed down by frequency-chirped counter-propagating laser light~\cite{Truppe2017a}. The slowest molecules are captured in a MOT~\cite{Williams2017} where the main laser drives the $A^{2}\Pi_{1/2}(v=0,J=1/2)\leftarrow X^{2}\Sigma^{+}(v=0,N=1)$ transition, with an intensity of $I$ and a detuning of $\delta$. The linewidth of this transition is $\Gamma$ = \SI{2\pi x 8.3}{\mega\hertz}. Three additional lasers repump population that leaks to the $v=1,2$ and 3 vibrational levels of the $X$ state. The MOT and magnetic trap share the same in-vacuum coils~\cite{Williams2017}, which produce an axial field gradient $B'$. Using $I=I_{\rm max}=\SI{400}{\milli\watt\per\square\centi\metre}$, $\delta = -0.75\Gamma$ and $B'=\SI{30}{\gauss\per\centi\metre}$, we routinely capture \num{2e4} molecules with a peak density of $n=\SI{6e5}{\per\cubic\centi\metre}$ and a temperature of $T=\SI{11}{\milli\kelvin}$.

Figure \ref{fig:structure_timing}(a) shows the energy levels most relevant to the present work, which we label $\ket{N, F, M_F}$. The ground rotational state, $N=0$, is split into two hyperfine components, $F=0$ and 1, while the first excited rotational state, $N=1$, is split by spin-rotation and hyperfine interactions into four components with $F\in\{1,0,1,2\}$. To address these, the MOT laser is tuned near $F=0$, the sideband of a \SI{48}{\mega\hertz} acousto-optic modulator (AOM) addresses the upper $F=1$ level and the sidebands of an electro-optic modulator (EOM) address the $F=2$ and lower $F=1$ levels. The light from the AOM and EOM have opposite circular polarizations, implementing a dual-frequency MOT~\cite{Tarbutt2015}. Since our previous work, we have changed the EOM frequency from \SI{74.5}{\mega\hertz} to \SI{70.5}{\mega\hertz}. This has increased the density by a factor of 4, mainly by increasing the MOT spring constant.

\begin{figure}[tb]
\includegraphics{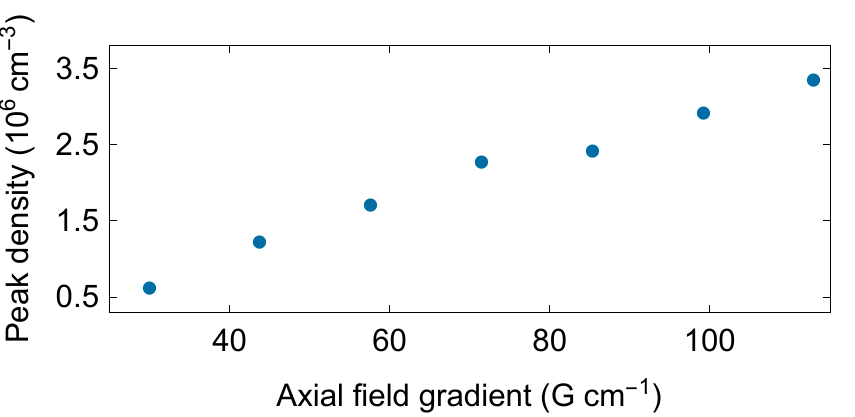}
\caption{Density as a function of the axial magnetic field gradient used for the compressed MOT.}
\label{fig:compressed_MOT}
\end{figure}

Figure \ref{fig:structure_timing}(b) illustrates the new control steps we implement and presents their timings. Each sequence begins with a fluorescence image taken at $t = \SI{40}{\milli\second}$, used to determine the number of molecules, $N_{\rm mol}$, in each MOT. We first compress the MOT by increasing $B'$ linearly between $t=40$ and \SI{50}{\milli\second}, and holding the higher $B'$ until $t = \SI{55}{\milli\second}$.  Figure~\ref{fig:compressed_MOT} shows $n$ as a function of $B'$ in the compressed MOT (cMOT). Increasing $B'$ to \SI{113}{\gauss\per\centi\meter}, increases $n$ to \SI{3.4e6}{\per\cubic\centi\meter}, a factor of 5.3 greater than in the standard MOT. If $N_{\rm mol}$ and $T$ are conserved in the compression, we expect $n \propto (B')^{3/2}$, resulting in a factor of 7.3 increase in density. We find that $N_{\rm mol}$ is conserved, but that $T$ increases, which explains the smaller observed factor. For all subsequent data, we use $B'=\SI{69}{\gauss\per\centi\metre}$ in the cMOT, giving $n\approx\SI{2e6}{\per\cubic\centi\metre}$.

Following the cMOT, we lower the temperature using a procedure similar to the one described previously~\cite{Truppe2017b}, and illustrated in Fig.~\ref{fig:structure_timing}(b). Between $t=55$ and \SI{59}{\milli\second}, $I$ is ramped down to $0.1I_{\rm max}$, where it is held until the MOT coils and laser are switched off at $t = \SI{63}{\milli\second}$. At $t = \SI{64}{\milli\second}$, the laser is re-enabled at full intensity but with $\delta=3\Gamma$, realizing a blue-detuned optical molasses. This configuration is held for \SI{6.5}{\milli\second}, then $I$ is stepped downwards to \SIlist{50;30;20}{\percent} of $I_{\rm max}$, with each value held for \SI{0.5}{\milli\second}. This procedure lowers $T$ to \SI{55}{\micro\kelvin}. The molasses hold time is far longer than the time taken to reach this temperature, which is less than \SI{1}{\milli\second}. This delay allows magnetic fields, caused by eddy currents induced when the coils switch off, to decay to $\lesssim$~\SI{1}{\milli\gauss}, which is needed for the subsequent microwave transfer step(s). The cloud expands slowly enough in the molasses that there is little loss in density during this period. 

\begin{figure*}
\includegraphics[width=\linewidth]{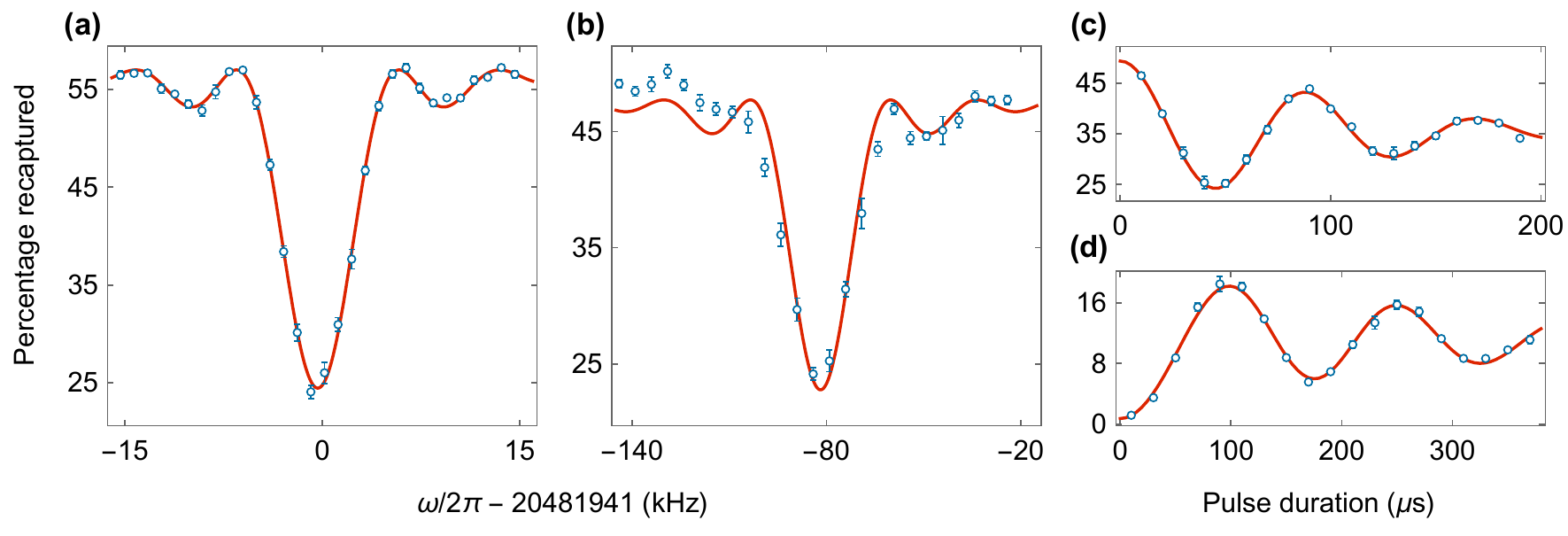}
\caption{Microwave transitions between rotational levels. (a,b) Depletion of $N=1$ population versus microwave frequency: (a) $\ket{1,0,0}\rightarrow\ket{0,1,0}$ transition driven by a  \SI{140}{\micro\second} $\pi$-pulse; (b) $\ket{1,0,0}\rightarrow\ket{0,1,1}$ transition driven by a  \SI{40}{\micro\second} $\pi$-pulse. (c) Rabi oscillations driven between $\ket{1,0,0}$ and $\ket{0,1,1}$. (d) Rabi oscillations between $\ket{0,1,1}$ and $\ket{1,2,2}$ after a push pulse has removed all molecules in $N=1$. Each point is an average of nine experiments, with an error bar that is the standard error of the set. Where not visible, error bars are smaller than the points. In (a,b), lines show fits to the Rabi lineshape for a $\pi$-pulse of the known duration, with the background, amplitude and central frequency as free parameters. The data have been corrected for a systematic frequency shift  of \SI{1.9}{\kilo\hertz} in (a) and \SI{4.7}{\kilo\hertz} in (b), arising from a small change in the microwave frequency when the pulse is applied. In (c,d), lines show fits to solutions of the optical Bloch equations with time varying detuning and integration over Rabi frequencies (see text).}
\label{fig:transitions}
\end{figure*} 

At this point, the population is distributed randomly amongst the 12 Zeeman sub-levels of ($v=0, N=1$) shown in Fig.~\ref{fig:structure_timing}(a), and the corresponding 12 levels in $v=1$. To purify the state distribution, we optically pump into $\ket{1,0,0}$. We do this by reducing $I$ to $0.01I_{\rm max}$ and increasing the rf power to the EOM so that the power in the carrier, which addresses $F=0$, is suppressed to $\leq$~\SI{1}{\percent} of the total. This transfers the population to $F=0$ with an efficiency $\epsilon_{\rm OP}=\SI{60}{\percent}$, limited by imperfect extinction of the carrier and off-resonant excitation by the other sidebands. It takes \SI{40}{\micro\second} for the population to reach its new steady state, though we pump for \SI{100}{\micro\second}.

With the majority of molecules now in a single state, we can transfer them to any selected Zeeman sub-level by driving microwave transitions between $N=0$ and $N=1$. The outputs of a two-channel microwave synthesizer are connected to a frequency doubler via a switch, and the doubled output is coupled to free space through a horn. The microwave field, tuned to a hyperfine component of the rotational transition, passes into the vacuum chamber along the same path as the slowing laser. Its polarization at the molecules is poorly defined due to reflections inside the chamber, and we find that $\Delta M_{F}=-1,0,1$ transitions can all be driven. Bias coils cancel the background magnetic field and apply a constant, uniform field of $\sim$~\SI{60}{\milli\gauss}, sufficient to resolve the Zeeman sub-levels, but small enough not to disrupt the molasses cooling.

Figure~\ref{fig:transitions}(a) shows the depletion of the $N=1$ population as a function of the microwave angular frequency, $\omega$, as it is scanned through the magnetically-insensitive $\ket{1,0,0}\rightarrow\ket{0,1,0}$ transition at  $\omega_0$. The microwave pulse has a duration of $\tau_{\mu 1}=\SI{140}{\micro\second}$, and the Rabi frequency is $\Omega = \pi/\tau_{\mu 1}$. Molecules transferred to $N=0$ are decoupled from the MOT light. Thus, we measure the number of molecules remaining in $N=1$ by turning the MOT back on and imaging the fluorescence after a time $\tau_{\rm hold}$, typically \SI{30}{\milli\second}. This number, divided by the initial number in the MOT, is the fraction recaptured. The line in Fig.~\ref{fig:transitions}(a) is a fit to the data using the model $y_{0}+A f(\Omega, \omega-\omega_0, \tau_{\mu 1})$, where $y_{0}$ is the fraction re-captured without the microwave pulse, $A$ is an amplitude, and $f$ is the usual Rabi lineshape for a two level system. We fix $\tau_{\mu 1}$ and $\Omega$, leaving $y_{0}$, $A$ and $\omega_0$ as free parameters. The fit is a good one, and gives $y_{0} = 0.57$, consistent with the MOT lifetime, and $A = -0.32$. The microwave transfer efficiency is $\epsilon_{\rm MW} = A/(y_{0}\epsilon_{\rm OP}) = \SI{94}{\percent}$. We infer that, in the relevant polarization, the microwave intensity at the molecules is \SI{64}{\nano\watt\per\square\centi\metre}. Figure~\ref{fig:transitions}(b) shows similar data for the magnetically-sensitive transition $\ket{1,0,0}\rightarrow\ket{0,1,1}$.  We drive a $\pi$-pulse with a shorter duration of $\tau_{\mu 1}=\SI{40}{\micro\second}$ in order to reduce the effects of magnetic field inhomogeneities and fluctuations. Their effects are still visible in the data, producing a slight broadening relative to the model, a poorer fit in the wings, and a lower efficiency of $\epsilon_{\rm MW} = \SI{87}{\percent}$. The inferred intensity in the relevant polarization is \SI{780}{\nano\watt\per\square\centi\metre}.

Figure \ref{fig:transitions}(c) shows Rabi oscillations on the magnetically-sensitive $\ket{1,0,0}\rightarrow \ket{0,1,1}$ transition. We measure the percentage recaptured versus $\tau_{\mu 1}$, with the microwave frequency on resonance and the microwave power held constant. To model these data we found it necessary to include two imperfections. The first relates to the microwave synthesizer, which we discovered has a transient frequency drift when switched. This frequency change is well modeled by $\omega(t') = \omega_{\infty} - \Delta\omega\,e^{-t'/\tau}$ where $\omega_{\infty}/(2\pi)$ is the frequency at long times, $t'$ is the time since the start of the pulse, $\Delta \omega/(2\pi)\approx \SI{7}{\kilo\hertz}$ is the total frequency change and $\tau \approx \SI{105}{\micro\second}$ is the timescale. This has no observable effect on the lineshapes in Fig.~\ref{fig:transitions}(a,b), but causes a slight frequency shift in the line centre, a noticeable chirp in the frequency of the Rabi oscillations and a slight reduction in their contrast. The second imperfection is due to  gradients of intensity and polarization produced by the standing wave component of the microwave field, and is the main reason for the gradual reduction in the contrast of the Rabi oscillations with increasing $\tau_{\mu 1}$. To model these effects, we first solve the two-level optical Bloch equations with the measured drift in frequency included. This gives a function $y_{0} + A g(\Omega, \omega_{\infty}-\omega_0, \tau_{\mu 1})$. We average this over a Gaussian distribution of Rabi frequencies with a width of $\Delta\Omega$. The solid line in Fig.~\ref{fig:transitions}(c) is a fit to this model, with $y_{0}$, $A$, $\omega_{\infty}-\omega_0$, $\Omega$ and $\Delta \Omega$ as free parameters. We find $\Delta \Omega / \Omega = 0.16$, which is reasonable since the distance from node to antinode of the standing wave component is \SI{3.5}{\milli\metre}, comparable to the size of the molecule cloud.

With most molecules now transferred to $N=0$, we push those remaining in $N=1$ out of the trap region by turning on the slowing light for \SI{1}{\milli\second}. This leaves a pure sample of molecules in a single state. We then either turn on the magnetic trap or apply a second microwave pulse, of duration $\tau_{\mu 2}$, to transfer back to a selected sub-level of $N=1$. Figure~\ref{fig:transitions}(d) shows Rabi oscillations on the $\ket{0,1,1}\rightarrow \ket{1,2,2}$ transition as $\tau_{\mu 2}$ is varied. The percentage recaptured is zero when $\tau_{\mu 2}=0$, showing that we indeed have a pure sample. The line is a fit using the same model described above and is seen to fit well. A $\pi$-pulse takes \SI{100}{\micro\second}, implying a microwave intensity in the appropriate polarization of \SI{42}{\nano\watt\per\square\centi\metre}. The efficiency of this second $\pi$-pulse is \SI{75}{\percent}. 

\begin{figure}
\includegraphics[width=\linewidth]{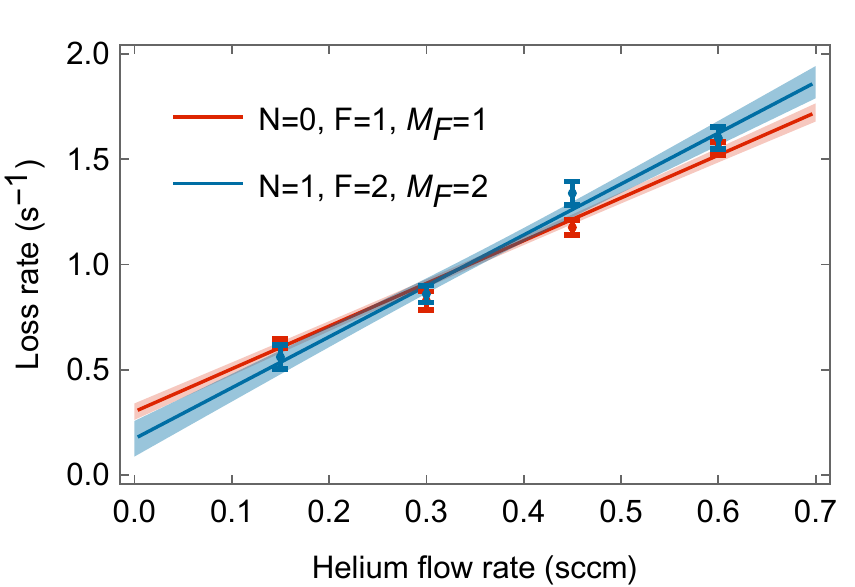}
\caption{Magnetic trap loss rate  versus helium flow rate through the buffer-gas source, for molecules in two different states, $\ket{0,1,1}$ (red) and $\ket{1,2,2}$ (blue). Lines are linear fits, and shaded regions are \SI{68}{\percent} confidence bands.}
\label{fig:trapping}
\end{figure}

With these procedures, we can trap molecules in any of the weak-field seeking states shown in Fig.~\ref{fig:structure_timing}(a). Here, we demonstrate trapping in the $\ket{0,1,1}$ and $\ket{1,1,2}$  states. We turn the magnetic trap on with $B' = \SI{30}{\gauss\per\centi\meter}$ after either the first or second microwave pulse, then detect the number remaining after a time $\tau_{\rm mag}$ by turning the MOT light on and imaging the fluorescence. When trapping molecules in $N=0$, which do not fluoresce in the MOT light, we transfer back to $N=1$ using the $\ket{0,1,1}\rightarrow \ket{1,2,2}$ transition prior to detection. Conveniently, this transition is magnetically-insensitive so can be driven while the molecules are magnetically trapped. Indeed, we observe Rabi oscillations on this transition, similar to those shown in Fig.~\ref{fig:transitions}(d), even when the molecules are trapped. The number of molecules in the trap fits well to a single-exponential decay, $N_{\rm mol}(\tau_{\rm mag}) = N_{\rm mol}(0) \exp(-R_{\rm loss} \tau_{\rm mag})$, which we attribute mainly to collisions with helium gas from the buffer-gas source. Figure~\ref{fig:trapping} shows the loss rate, $R_{\rm loss}$, as a function of the helium flow rate, for molecules in each of the two states, showing a linear dependence in both cases. The gradients are \SIlist{2.03+-0.08;2.42+-0.16}{\per\second\per\sccm} for the $\ket{0,1,1}$ and $\ket{1,1,2}$ states respectively, differing by $~\SI{2.2}{\sig}$.  Extrapolating to zero flow, the loss rates for the two states are \SIlist{0.30+-0.03;0.17+-0.06}{\per\second}, differing by $~\SI{1.9}{\sig}$.  These rates are close to the loss rate due to vibrational excitation by room temperature blackbody radiation~\cite{Buhmann2008}, which is \SI{0.22}{\per\second} for all the states shown in Fig.~\ref{fig:structure_timing}(a).

To investigate whether the molecules are heated in the magnetic trap, we have measured their temperature prior to trapping and for various $\tau_{\rm mag}$. Before turning on the trap, the radial, axial, and geometric mean temperatures are $\{T_{\rho},T_{z}, T\} = \{52(12), 69(12), 57(9)\}$~\SI{}{\micro\kelvin}. At $\tau_{\rm mag} = \SI{40}{\milli\second}$ we measure $\{56(12), 131(12), 74(11)\}$~\SI{}{\micro\kelvin}, while at $\tau_{\rm mag} = \SI{500}{\milli\second}$ the values are $\{47(12), 123(12), 65(11)\}$~\SI{}{\micro\kelvin}. These measurements show that trap loading results in modest heating, though only in the axial direction, and that the heating rate in the trap is consistent with zero and has an upper limit of \SI{37}{\micro\kelvin\per\second}. We load about \SI{5e3} molecules into the trap, and the cloud has radial and axial rms radii of $\sigma_{\rho} = \SI{1.37+-0.01}{\milli\metre}$ and $\sigma_{z} = \SI{1.44+-0.02}{\milli\metre}$. The density is \SI{1.2e5}{\per\cubic\centi\metre}, and could be increased by increasing $B'$. The phase-space density is \num{2.6e-12}.

In summary, we have compressed our MOT to increase the density of molecules, demonstrated coherent control of their rotational, hyperfine and magnetic states, and transferred them to a conservative trap. The entire sequence takes only \SI{75}{\milli\second}. We have measured the trap loss rate for two selected states, which is a prototype measurement for future experiments studying state-selective elastic and inelastic collisions between co-trapped ultracold atoms and molecules~\cite{Parazzoli2011}. Our demonstrations of long lifetimes and low heating rates in a magnetic trap are important for reaching lower temperatures by sympathetic cooling~\cite{Tokunaga2011,Lim2015}. The magnetically trapped molecules can now easily be transported to experiments more conveniently located away from the MOT region~\cite{Lewandowski2003}. The quantum state control we demonstrate is required for measurements with ultracold molecules that test fundamental physics~\cite{Tarbutt2013, Hunter2012, Cahn2014,Cheng2016}. With sufficient control, quantum information can be stored within hyperfine states and the dipole-dipole interaction turned on when needed for information processing~\cite{Andre2006}. Controlled microwave-induced dipoles are crucial for simulating spin Hamiltonians~\cite{Micheli2006, Barnett2006, Gorshkov2011, Yan2013}, studying topological superfluids~\cite{Cooper2009} and enhancing evaporative cooling~\cite{Avdeenkov2012}. For a realistic spacing of \SI{0.5}{\micro\metre}, the dipole-dipole interaction energy is a few kHz, comparable to the resolution we achieve. Thus, our work demonstrates many of the key capabilities needed for the applications of laser-cooled molecules.

We are grateful to J. Dyne, G. Marinaro and V. Gerulis for technical assistance. The research has received funding from
EPSRC under grants EP/M027716/1, and
EP/P01058X/1, and from the European Research Coun-
cil under the European Union's Seventh Framework
Programme (FP7/2007-2013) / ERC grant agreement
320789.  Data underlying this article can be accessed from Zenodo~\footnote{\lowercase{https://doi.org/10.5281/zenodo.1035877}} and may be used under the Creative Commons CCZero license.

H. J. Williams and L. Caldwell contributed equally to this work.

\bibliography{references}

\end{document}